\documentclass[twocolumn,aps,prd,10pt,superscriptaddress,longbibliography]{revtex4-1}
\usepackage{amsmath}
\usepackage{bm}
\usepackage{soul}
\usepackage{graphicx}
\usepackage{multirow}
\usepackage[usenames,dvipsnames,svgnames]{xcolor}
\usepackage{hyperref}
\hypersetup{
pdfnewwindow=true,      
colorlinks=true,        
linkcolor=Blue,         
citecolor=Blue,         
filecolor=Blue,         
urlcolor=Blue           
}

\begin{document}

\title{Exactly equivalent thermal conductivity in finite systems from  equilibrium and nonequilibrium molecular dynamics simulations}

\author{Haikuan Dong}
\affiliation{Beijing Advanced Innovation Center for Materials Genome Engineering, Corrosion and Protection Center, University of Science and Technology Beijing, Beijing, 100083, China}
\affiliation{College of Physical Science and Technology, Bohai University, Jinzhou, 121013, China}
\author{Zheyong Fan}
\email{brucenju@gmail.com}
\affiliation{College of Physical Science and Technology, Bohai University, Jinzhou, 121013, China}
\author{Ping Qian}
\email{qianping@ustb.edu.cn}
\affiliation{Beijing Advanced Innovation Center for Materials Genome Engineering, Department of Physics, University of Science and Technology Beijing, Beijing 100083, China}
\author{Yanjing Su}
\email{yjsu@ustb.edu.cn}
\affiliation{Beijing Advanced Innovation Center for Materials Genome Engineering, Corrosion and Protection Center, University of Science and Technology Beijing, Beijing, 100083, China}

\begin{abstract}
In a previous paper [Physical Review B \textbf{103}, 035417 (2021)], we showed that the equilibrium molecular dynamics (EMD) method can be used to compute the apparent thermal conductivity of finite systems. It has been shown that the apparent thermal conductivity from EMD for a system with domain length $2L$ is equal to that from nonequilibrium molecular dynamics (NEMD) for a system with domain length $L$. Taking monolayer silicence with an accurate machine learning potential as an example, here we show that the thermal conductivity values from EMD and NEMD agree for the same domain length if the NEMD is applied with periodic boundary conditions in the transport direction. Our results thus establish an exact equivalence between EMD and NEMD for thermal conductivity calculations.
\end{abstract}
\maketitle

\section{Introduction}
Molecular dynamics (MD) is one of the most important methods for studying heat transport at the nanoscale. The equilibrium MD (EMD) method based on the Green-Kubo relation  \cite{green1954jcp,kubo1957jpsj} and the nonequilibrium MD (NEMD) method based on a direct simulation of steady-state heat flow  \cite{ikeshoji1994mp,muller-plathe1997jcp,jund1999prb,shiomi2014arht} are the two canonical methods for computing the lattice thermal conductivity in various materials. There have been extensive comparisons between the two methods for diffusive phonon transport  \cite{schelling2002prb,sellan2010prb,he2012pccp,howell2012jcp,dong2018prb} and it is now generally accepted \cite{gu2021jap} that the two methods give consistent results in the diffusive limit.

While it has been believed that the EMD method can only be used to compute the thermal conductivity in the diffusive limit, we have recently shown \cite{dong2021prb} that it can also be used to compute the apparent thermal conductivity (also called the effective thermal conductivity) of finite systems. For a finite system with nonperiodic boundary conditions applied to the transport direction, the running thermal conductivity (RTC) based on the Green-Kubo relation will eventually converges to zero in the limit of infinite-time, but it has been argued \cite{dong2021prb} that the maximum value of the RTC can be interpreted as the apparent thermal conductivity of the finite system. However, it was found that the apparent thermal conductivity as computed from the EMD method for a system with domain length $2L$ is equal to that as computed from the NEMD method for a system with domain length $L$. This difference between the domain length is puzzling and in this work we show that it is related to the choice of simulation setups in the NEMD method:  if the NEMD is performed with periodic boundary conditions, instead of fixed boundary conditions, in the transport direction, the thermal conductivity values from EMD and NEMD will agree for the same domain length. To be specific, our results are based on a case study of two-dimensional (2D) monolayer silicene using a recently developed machine learning potential \cite{fan2021prb} that can accurately capture the phonon properties of this material.

\section{Models and Methods}

\subsection{The EMD simulation setup}

\begin{figure}[htb]
\begin{center}
\includegraphics[width=\columnwidth]{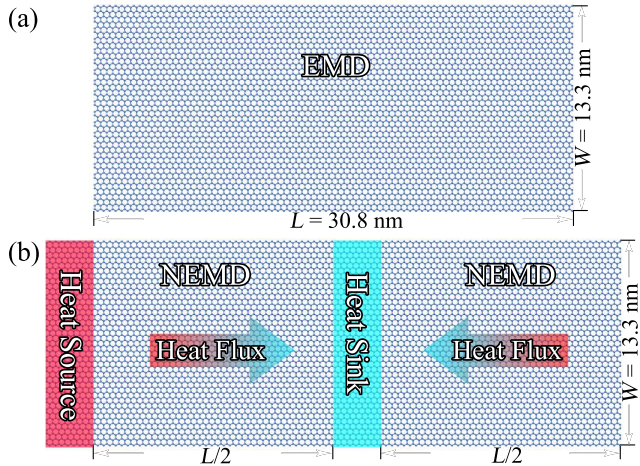}
\caption{Schematic illustration of the EMD and NEMD simulation setups. (a) EMD simulation setup with open boundary conditions in the transport direction (horizontal) with domain length $L$. (b) NEMD simulation setup with periodic boundary conditions in the transport direction (horizontal). In NEMD, the heat source and sink are separated by $L/2$. In both EMD and NEMD, the transverse direction (vertical) has periodic boundary conditions and width $W$.}
\label{fig01}
\end{center}
\end{figure}

The EMD simulation setup considered in this paper is schematically illustrated in Fig.~\ref{fig01}(a). To be specific, we align the zigzag direction of silicene in $x$ and the armchair direction in $y$ and study heat transport in the $x$ direction. Periodic boundary conditions are applied to the $y$ direction. In the $x$ direction, we use the free (open) boundary conditions to model a finite domain length $L$.

In the EMD method, after thermal equilibration, the system is evolved in the microcanonical ensemble and the equilibrium heat current \cite{fan2015prb}
\begin{equation}
    \vec{J}= \sum_i\sum_j \vec{r}_{ij} \frac{\partial U_j}{\partial \vec{r}_{ji}} \cdot \vec{v}_i
\end{equation}
is sampled. Here, $U_j$ is the site potential of atom $j$, $\vec{v}_i$ is the velocity of atom $i$, and $\vec{r}_{ij}=-\vec{r}_{ji} = \vec{r}_j - \vec{r}_i$ is the relative position from atom $i$ to atom $j$. From the sampled heat current one can then calculate the heat current autocorrelation function (HCACF) $\langle J_{x}(0) J_{x}(t)\rangle$ and the RTC $\kappa(\tau)$ through the following Green-Kubo relation \cite{green1954jcp,kubo1957jpsj}:
\begin{equation}
\kappa(\tau) = \frac{1}{k_{\rm B} T^2 V}\int_0^{\tau} \langle J_{x}(0) J_{x}(t) \rangle dt.
\end{equation}
Here, $k_{\rm B}$ is Boltzmann's constant, $T$ is the system temperature, and $V$ is the system volume. 

\subsection{The NEMD simulation setup}

\begin{figure}[tb]
\begin{center}
\includegraphics[width=1\columnwidth]{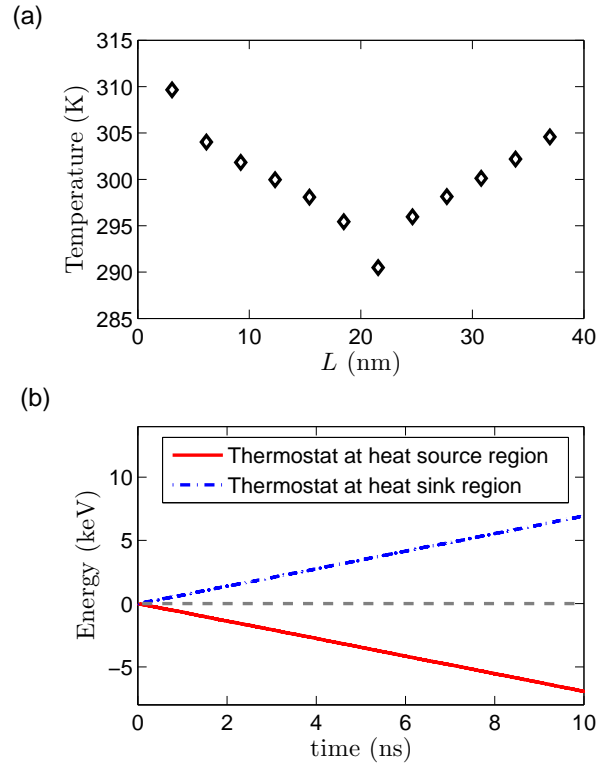}
\caption{ (a) Steady-state temperature profile and (b) energies of the thermostats coupled to the heat source and sink regions as a function of time in the NEMD simulation as shown in Fig. \ref{fig01}(b). The energy of the high-temperature thermostat  decreases as it releases energy to the heat source region; the energy of the low-temperature thermostat  increases as it absorbs energy from the heat sink region.}
\label{fig02}
\end{center}
\end{figure}

Figure \ref{fig01}(b) shows the NEMD simulation setup with periodic boundary conditions in the transport direction ($x$ direction). To be consistent with the EMD simulations, periodic boundary conditions are applied to the $y$ direction. In the $x$ direction, there are two thermostatted regions that are separated by $L/2$, serving as a heat source (with a higher temperature $T+\Delta T/2$) and a heat sink (with a lower temperature $T-\Delta T/2$). The total length of the parts without thermostat is thus $L$, which is the same as in the case of EMD. With a temperature difference of $\Delta T=20$ K, a typical temperature profile in the NEMD simulations is shown in Fig. \ref{fig02}(a).

In the NEMD method, the heat source region will absorb energy from the high-temperature thermostat at a given rate $dE/dt$ when steady state is achieved. According to energy conservation, the heat sink region will simultaneously release energy to the low-temperature thermostat at the same rate, as shown in Fig. \ref{fig02}(b). Denoting the cross-sectional area as $A$, the heat flux in one direction shown in Fig. \ref{fig01}(b) is $dE/dt/(2A)$, where the factor $2$ accounts for the fact that the heat flows from the heat source region to the heat sink region \textit{via} two opposite directions. Then, according to Fourier's law, the apparent (effective) thermal conductivity $\kappa(L)$ of the finite system with domain length $L$ is 
\begin{equation}
\label{equation:nemd}
\kappa(L) = \frac{dE/dt/(2A)}{\Delta T / (L/2)},
\end{equation}
Here the temperature gradient is taken as $\Delta T/(L/2)$, following the established practice  \cite{li2019jcp,hu2020prb}. 

\subsection{Other MD simulation details}

In both EMD and NEMD simulations, the width in the transverse direction is chosen as $W=13.3$ nm. The thickness of the monolayer silicene was chosen as the conventional value of $0.42$ nm \cite{zhang2014prb,dong2018prb,zhang2019jap,fan2021prb}. We consider the following domain lengths in the transport direction: $L$ = $7.7$, $15.4$, $30.8$, $61.6$, and $123.2$ nm.
 
For both EMD and NEMD simulations, we first equilibrate the system in the isothermal ensemble at $300$ K and zero pressure for $0.1$ ns. Then, in the EMD simulations, we evolve the system in the microcanonical ensemble for $5$ ns. In the NEMD simulations, we generate the nonequilibrium heat current using the local Langevin thermostats (with a coupling time of $0.1$ ps) for $10$ ns and use the second half of the data to compute the steady-state heat flux. We perform $30$ and $3$ independent runs for each domain length in the EMD and NEMD simulations, respectively. All the MD simulations are performed by using the open-source \textsc{gpumd} package \cite{fan2013cpc,fan2017cpc}. A time step of $1$ fs is used in all the cases. 

\subsection{The NEP machine learning potential for silicene}

We use the recently developed neuroevolution potential (NEP) \cite{fan2021prb}, which is a machine learning potential framework that can achieve high accuracy and low cost simultaneously. 

The training data were obtained using an active learning scheme as implemented in the \textsc{mlip} package \cite{Novikov2021}. We have considered temperatures ranging from $100$ to $900$ K and biaxial in-plane strains ranging from $-1\%$ to $1\%$. There are $914$ structures with $54840$ atoms in total. More details on the training data and the relevant hyperparameters in the NEP potential have been presented in Ref. \cite{fan2021prb}. The trained NEP potential as well as other inputs and outputs are freely available \cite{nep-data}.

\begin{figure}[htb]
\begin{center}
\includegraphics[width=\columnwidth]{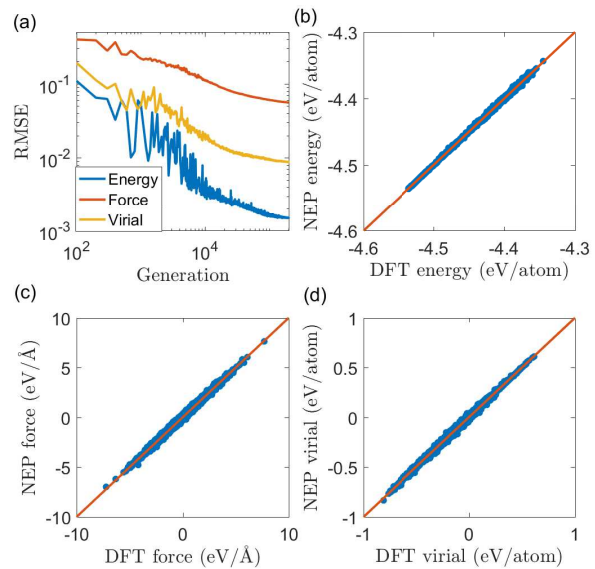}
\caption{(a) Evolution of the energy, force, and virial RMSEs during the NEP training process. (b)-(d) The predicted energy, force, and virial values from the trained NEP potential as compared to the reference values from quantum-mechanical DFT calculations.} 
\label{fig03}
\end{center}
\end{figure}

Figure \ref{fig03}(a) shows the evolution of the energy, force and virial root mean square errors (RMSEs) as a function of the number of generations during the training process. The predicted energy, force, and virial values from the trained NEP potential are compared with the reference values from quantum-mechanical density functional theory (DFT) calculations in Figs. \ref{fig03}(b)-\ref{fig03}(d). The RMSEs for the energy, force components, and virial components are $1.5$ meV/atom, $56$ meV/\AA, and $8.8$ meV/atom, respectively. These RMSEs are comparable to those obtained from other popular machine learning potentials but the NEP potential we use is much more computationally efficient \cite{fan2021prb}. For example, our NEP potential can achieve a speed of $1.3\times 10^7$ atom-step per second in MD simulations using a single Nvidia Tesla V100 GPU, while the moment tensor potential as implemented in \textsc{mlip} can only achieve a speed of $10^5$ atom-step per second using 72 Intel Xeon-Gold 6240 CPU cores. The efficiency of the NEP potential allows for doing large-scale and long-time MD simulations with a reasonable amount of computational resources.

\section{Results and discussion}

\begin{figure*}[htb]
\begin{center}
\includegraphics[width=1.8\columnwidth]{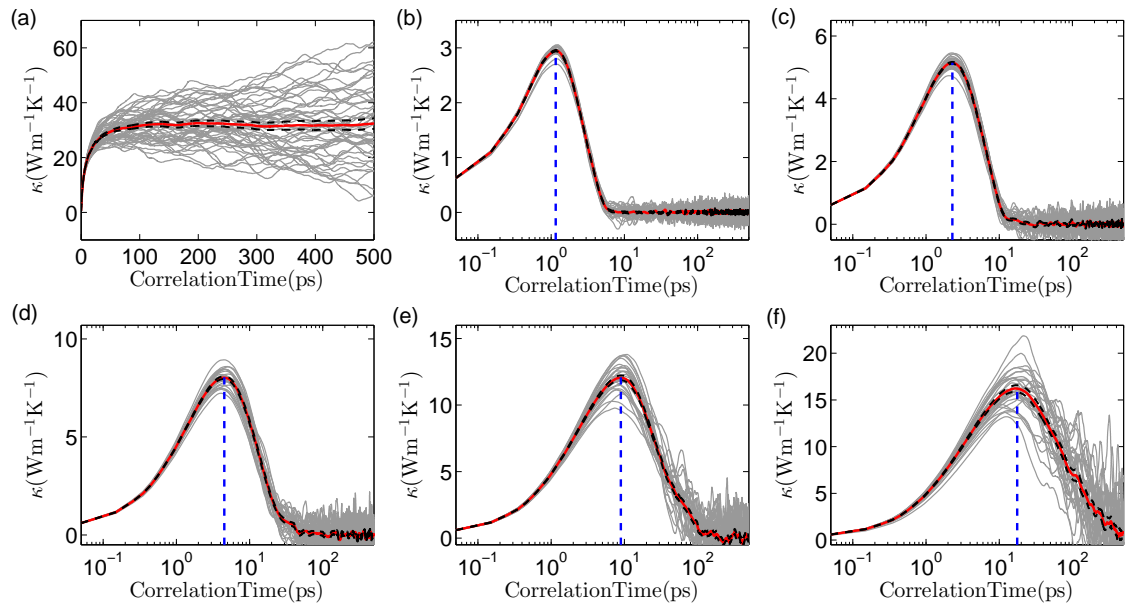}
\caption{RTC $\kappa(\tau)$ as a function of the upper integration limit of the correlation time $\tau$ from EMD simulations with (a) periodic boundary conditions and  (b)-(f) open boundary conditions in the transport direction. The domain lengths in (b)-(f) are $L$ = $7.7$, $15.4$, $30.8$, $61.6$, and $123.2$ nm. In each panel, the red thick line represents the mean value from the independent runs (the gray thin lines) and the black dashed lines indicate the error bounds. The blue dashed vertical lines in (b)-(f) correspond to the time $\tau_{\rm max}$ at which $\kappa(\tau)$ reaches a maximum.} 
\label{fig04}
\end{center}
\end{figure*}

Figure \ref{fig04}(a) shows the RTC $\kappa(\tau)$ of silicene as a function of the upper limit of the correlation time $\tau$ in the Green-Kubo integral. Here, periodic boundary conditions are applied to the transport direction and the simulated system can be considered as an infinite 2D system. In this case, the RTC will converge to a finite value which can be taken as the thermal conductivity of 2D silicene in the diffusive limit. We have used a sufficiently large domain size of $38.7$ $\times$ $40.2$ nm$^2$ with $24000$ atoms and the converged thermal conductivity is $31.9$ $\pm$ $1.4$ $\rm Wm^{-1}K^{-1}$, Which is consistent with the previous value ($33.7$ $\pm$ $0.6$ $\rm Wm^{-1}K^{-1}$) from the same NEP potential reported in Ref. \cite{fan2021prb} as well as that ($32.4$ $\pm$ $2.9$ $\rm Wm^{-1}K^{-1}$) from a Gaussian approximation machine learning potential reported in Ref. \cite{zhang2019jap}.

While the RTC converges to a finite value when periodic boundary conditions are applied in the transport direction, it always converges to zero when nonperiodic (open or fixed) boundary conditions are applied in the transport direction. This can be seen in Figs \ref{fig04}(b)-\ref{fig04}(f) for all the domain lengths considered in this paper. Despite of the zero value in the infinite-time limit, the RTC in these finite systems exhibits a clear maximum value $\kappa^{\rm EMD}_{\rm max}$ at a particular correlation time $\tau_{\rm max}$. In our previous work \cite{dong2021prb}, this maximum value has been interpreted as the apparent thermal conductivity in the finite systems. Similar to Ref. \cite{dong2021prb}, we find that $\tau_{\rm max}$ is almost a linear function of the domain length $L$, $L \approx v_{\rm g} t_{\rm max}$, where the proportionality coefficient $v_{\rm g}$ can be interpreted as an effective group velocity of the phonons, which is fitted to be about $7$ km s$^{-1}$. 

\begin{table}[htb]
\caption{Thermal conductivity data for finite systems with length ($L$) from the EMD ($\kappa^{\rm EMD}_{\rm max}$) and NEMD ($\kappa^{\rm NEMD}$) simulations.}
\label{table:tab1}
\begin{tabular}{rrrrrr}
\hline
\hline
$L$  (nm)  & $\kappa^{\rm EMD}_{\rm max}$ ($\rm Wm^{-1}K^{-1}$) & $\kappa^{\rm NEMD}$ ($\rm Wm^{-1}K^{-1}$)\\
\hline
7.7   & 2.95$\pm$0.01                 & 2.91$\pm$0.01           \\
15.4  & 5.15$\pm$0.03                 & 5.12$\pm$0.05           \\
30.8  & 8.04$\pm$0.07                 & 7.94$\pm$0.09           \\
61.6  & 12.12$\pm$0.17                & 11.61$\pm$0.12          \\
123.2 & 16.54$\pm$0.33                & 16.03$\pm$0.21          \\
\hline
\hline
\end{tabular}
\end{table}

\begin{figure}[htb]
\begin{center}
\includegraphics[width=\columnwidth]{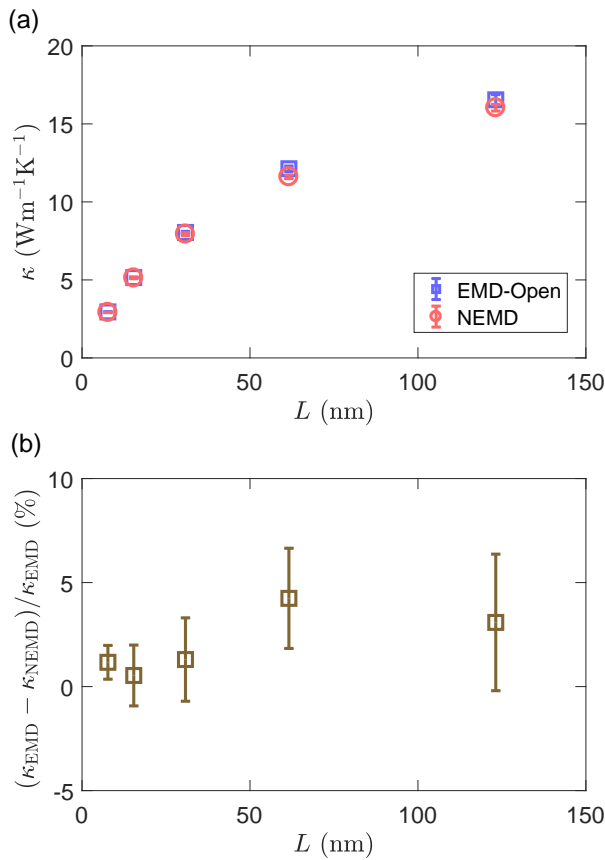}
\caption{(a) The maximum thermal conductivity $\kappa_{\rm max}$ from EMD simulations with open boundary conditions in the transport direction and the apparent thermal conductivity from NEMD simulations as a function of the domain length $L$ for monolayer silicene at 300 K and zero pressure. (b) The relative difference of thermal conductivity between EMD and NEMD as a function of the domain length $L$.} 
\label{fig05}
\end{center}
\end{figure}

The maximum RTC in EMD simulations with different $L$ are listed in Table \ref{table:tab1}. In this table, we also list the thermal conductivity values from NEMD simulations as calculated using Eq. (\ref{equation:nemd}). The data are also plotted in Fig. \ref{fig05}(a). It is clear that the thermal conductivity values from EMD and NEMD agree with each other excellently. As can be seen from Fig. \ref{fig05}(b), the relative difference of thermal conductivity between EMD and NEMD is less than $5\%$, which is comparable to the statistical uncertainty in the MD simulations. Mathematically, we have
\begin{equation}
\label{equation:emd_nemd}
\kappa^{\rm EMD}_{\rm max}(L) = \kappa^{\rm NEMD}(L).
\end{equation}
Here we emphasize that the EMD results are obtained by using nonperiodic boundary conditions in the transport direction. We have used the open boundary conditions, but similar results can also be obtained by using fixed boundary conditions in the EMD simulations. 

Different from Ref. \cite{dong2021prb}, where a relation $\kappa^{\rm EMD}_{\rm max}(2L) = \kappa^{\rm NEMD}(L)$ has been found, we do not have a difference of domain length between EMD and NEMD simulations in Eq. (\ref{equation:emd_nemd}). This means that the EMD and NEMD simulation setups in Fig. \ref{fig01} are directly comparable. Indeed, in both setups, the average phonon mean free paths related to phonon-boundary scattering are the same: in NEMD simulations, it is the separation of the two thermostats, which is $L/2$; in EMD simulations, it is the average path length between any point in the domain and a boundary, which is also $L/2$. It is this equivalence that gives rise to the same apparent thermal conductivity for a given domain length from the two methods. 

While the two methods are equivalent in computing the apparent thermal conductivity in finite systems, they are not the methods of choice for computing the diffusive thermal conductivity in the infinite-length limit. Indeed, when $L=123.2$ nm, the apparent thermal conductivity is still only $16.03 \pm 0.21$ $\rm Wm^{-1}K^{-1}$ from NEMD, which is only about one half of the diffusive thermal conductivity. However, at the nanoscale, one frequently needs to consider finite systems and the equivalence between EMD and NEMD demonstrated in this paper will be useful as one can choose the appropriate method that is more suitable for a particular application. One application that is more suitable for the EMD method is the determination of the apparent thermal conductivity of some zero-dimensional systems with irregular shapes, such as nanoparticles, in which case the application of the NEMD method is not very straightforward \cite{Matsubara2020drm}.

\section{Summary and Conclusions}

In summary, we have calculated the apparent thermal conductivity of monolayer silicene with finite domain lengths based on an accurate NEP machine learning potential using EMD and NEMD methods. For a finite system of simulation domain length $L$, the maximum thermal conductivity $\kappa^{\rm EMD}_{\rm max}(L)$ from EMD simulations with open boundary conditions corresponds exactly to the apparent thermal conductivity $\kappa^{\rm NEMD}(L)$ from NEMD simulations with periodic boundary conditions in the transport direction. The EMD method with nonperiodic boundary conditions in the transport direction could be used in situations where the NEMD method is not straightforward to apply, such as characterizing the apparent thermal conductivity of finite nanoparticles. 

\section*{Acknowledgments}
This work was supported by the National Key Research and Development Program of China under Grant Nos. 2021YFB3802100 and 2018YFB0704300, the National Natural Science Foundation of China under Grant No. 11974059, the Science Foundation from Education Department of Liaoning Province under Grant No. LQ2020008. We acknowledge the computational resources provided by High Performance Computing Platform of Beijing Advanced Innovation Center for Materials Genome Engineering.

\end{document}